# Superconducting Nanowires as Nonlinear Inductive Elements for Qubits

Jaseung Ku[(1,*)], Vladimir Manucharyan[(2,*)], Alexey Bezryadin[(1)]

(1) University of Illinois at Urbana-Champaign

1110 W. Green Street, Urbana, IL 61801

(2) Yale University

15 Prospect Street, New Haven, CT 06511

**Abstract**

*We report microwave transmission measurements of superconducting Fabry-Perot resonators (SFPR), having a superconducting nanowire placed at a supercurrent antinode. As the plasma oscillation is excited, the supercurrent is forced to flow through the nanowire. The microwave transmission of the resonator-nanowire device shows a nonlinear resonance behavior, significantly dependent on the amplitude of the supercurrent oscillation. We show that such amplitude-dependent response is due to the nonlinearity of the current-phase relationship (CPR) of the nanowire. The results are explained within a nonlinear oscillator model of the Duffing oscillator, in which the nanowire acts as a purely inductive element, in the limit of low temperatures and low amplitudes. The low quality factor sample exhibits a "crater" at the resonance peak at higher driving power, which is due to dissipation. We observe a hysteretic bifurcation behavior of the transmission response to frequency sweep in a sample with a higher quality factor. The Duffing model is used to explain the Duffing bistability diagram. We also propose a concept of a nanowire-based qubit that relies on the current dependence of the kinetic inductance of a superconducting nanowire.*

## INTRODUCTION

Macroscopic quantum mechanics is one of the most exciting branches of modern physics, which, among other things, holds a promise for quantum computation applications. The program of studying macroscopic quantum phenomena, such as laboratory versions of "Schrödinger's Cat", was initiated by Leggett [1,2,3,4,5] in 70' and 80'. Perhaps the most important step in the confirmation of macroscopic quantum mechanics was the experimental observation of quantum behavior in superconducting micron-scale device reported by Martinis, Devoret, and Clark [6]. These experiments employed microwave probing of the discrete energy spectrum and macroscopic quantum tunneling (MQT) in a superconducting device. These developments led to the concept of a quantum computation [7,8,9,10,11], in which electronic devices, carrying bits of information, can exist in a quantum superposition of macroscopically distinct states and therefore can act as quantum bits (qubits).

One promising approach for constructing a superconducting qubit is to use an inductive element whose kinetic inductance ( $L$ ) depends on the magnitude of the supercurrent. If such an element is included into a superconducting LC-circuit, the current-dependence of the inductance makes the resonator anharmonic. The superconducting resonators, as any resonator in general, possesses a discrete energy spectrum. If the inductance shows a sufficient current dependence, one can make a nonlinear superconducting resonator. The goal is to make a resonator such that the energy difference between the ground and the first excited state is significantly different (i.e. larger than the width of the levels) than the difference between the first and the second excited states. Such resonator can be used as a qubit, since it can be manipulated between two bottom energy levels only, without ever exciting the third level.

We propose and test the possibility of using superconducting nanowires [12] as nonlinear inductive elements. In the future they can be expected to replace the usual superconductor-insulator-superconductor (SIS) Josephson junction [13] in qubits. The reason for moving away from SIS junctions is that they possess an insulating barrier controlling the critical current. A small number of impurities in the junction barrier can profoundly affect its physical properties [14] and cause decoherence [15]. On the other hand, the nanowires do not have an insulating barrier since their critical current is controlled by the wire diameter. Fine-tuning of the critical current can be achieved by connecting two nanowires in parallel and applying a weak magnetic field perpendicular to the formed loop [16]. Moreover, the superconducting Dayem bridge was shown theoretically to provide sufficient anharmonicity for quantum bits [17].

The current-phase relationship (CPR) $I(\delta)$ of an SIS junction is $I = I_0 \sin(\delta)$, while that of a short superconducting nanowire is $I = 2I_0 \cos(\delta/2)\text{arcth}[\sin(\delta/2)]$ [18], where $\delta$ is the gauge invariant phase difference across the wire. The kinetic inductance of a superconducting weak link, in general, is $L_{eff} = (h/4\pi e)(dI/d\delta)^{-1}$. It depends on $\delta$ [18] and, in turn, on the supercurrent flowing through the wire, hence explaining the term "non-linear inductance" (here *h* is the Planck's constant and *e* is the electronic charge). Such dependence is due to the fact that as the supercurrent approaches the critical depairing current, the superfluid density is suppressed and the response to a phase difference is changed. At $k_BT<<\Delta$ and in the absence of out-of-equilibrium Bogoliubov quasiparticles (BQ), this inductance is expected to be dissipationless, which is exactly what is required for coherent qubit operation. Yet, it is not clear *a priori* whether the suppression of the superfluid density in the wire by a high supercurrent would produce quasiparticles or not (such suppression is required exactly for the purpose of changing the kinetic inductance of the wire). Our experiments show that, for MoGe superconducting nanowires, there is a range of bias currents in which there is no additional dissipation while nonlinearity of the kinetic inductance is significant.

The nonlinear inductance of nanowires was probed by placing a nanowire under investigation into a superconducting coplanar waveguide resonator (i.e. a type of Fabry-Perot (FP) resonator, to be described in detail below), at a location where the microwave field imposes a sinusoidal time-dependent supercurrent and, correspondingly, a time-dependent phase difference $\delta$ [19]. Two devices are reported (S1 and S2). In such devices the oscillating supercurrent with full amplitude is forced, due to geometry of the device, to flow through the nanowire. Thus the resonance frequency and the quality factor of the device depend on the kinetic inductance of the nanowire and the dissipation occurring in the nanowire. The transmission amplitude of the resonator was measured and a sequence of resonance peaks was observed. One important result was the observation of the current-dependent kinetic inductance of the nanowire: As the driving amplitude was increased, the resonance peak shifted to lower frequencies. With the sample studied the shift was significant, i.e. of the order of the peak width. As this shift happens the quality factor does not change significantly. This fact indicates that BQs are not generated when the nonlinearity is present. Thus the nanowires could be potentially useful for the implementation of qubits, provided that the current remains sufficiently lower than the critical depairing current. Further increase of the driving power leads to a hysteric bifurcation of the transmission amplitude (on samples with higher quality factors), which can be explained by a Duffing model [20]. Thus we were able to construct and compare to the theory the Duffing bistability diagram. It is also found that regimes of stronger dissipation (such a "crater" shape of the resonance peak or a flattening of the resonance peak) can also be achieved if the amplitude of the supercurrent approaches closely the depairing current. Thus, for potential qubit applications, one would need to stay in the intermediate amplitude regime in which the nonlinearity is present but the dissipation due to the wire is still negligible. We demonstrate that such regime does exist.

In order to argue that a thin wire would be sensitive enough to provide nonlinearity to the resonator even in the quantum regime, i.e. when the number of photons in the system is of order unity, we stress the following fact: Even a single photon can induce a supercurrent that exceeds the critical current of a typical nanowire. To demonstrate this we analyze zero-point fluctuations and the average energy stored in the resonator. We model our resonator with a nanowire as a quantum harmonic oscillator, namely an LC-circuit. We apply the principle of equipartition of kinetic and potential energies in a harmonic oscillator, i.e. use the condition $\left\langle LI^2/2 \right\rangle = \hbar\omega_0/4$, where $L$ is the effective total inductance of the resonator, typically of the order of 1 nH, and $I$ is the supercurrent at the antinode of the resonator, and $\omega_0$ is the fundamental resonance frequency. It shows, for example, that in a $\omega_0/2\pi$ =10 GHz resonator made of a coplanar waveguide with characteristic impedance $Z_0 = 50\Omega$, even zero-point fluctuations generate a root mean square (RMS) current of $I_{rms} = \sqrt{\hbar\omega_0/2L} \approx \omega_0\sqrt{\hbar/2Z_0} \cong 60\,\text{nA}$, where the usual expressions $Z_0 \approx \sqrt{L/C}$ and $\omega_0 = 1/\sqrt{LC}$ have been assumed (here $C$ is the effective total capacitance). This value is comparable to the observed values of the critical currents of some MoGe nanowires. For example, a critical current of ~200 nA was previously reported [21]. Note that a precise in situ tuning of the critical current of a nanowire is possible by voltage pulsing [22]. If, on the other hand, a single photon is introduced into the resonator, the supercurrent RMS value should be $\sqrt{3}$ time larger than that of the zero-point fluctuations. Thus it appears quite practical to achieve a situation where by placing just a few photons in the resonator the supercurrent would approach the critical current. Thus the inductance of the wire should be significantly different in the situations where there is one photon in the resonator and two photons in the resonator, simply because its kinetic inductance changes with the supercurrent, due to the suppression of the superfluid density. Thus the desired level of nonlinearity and quantum single-photon manipulations of the device appears plausible.

**a)** **b)**

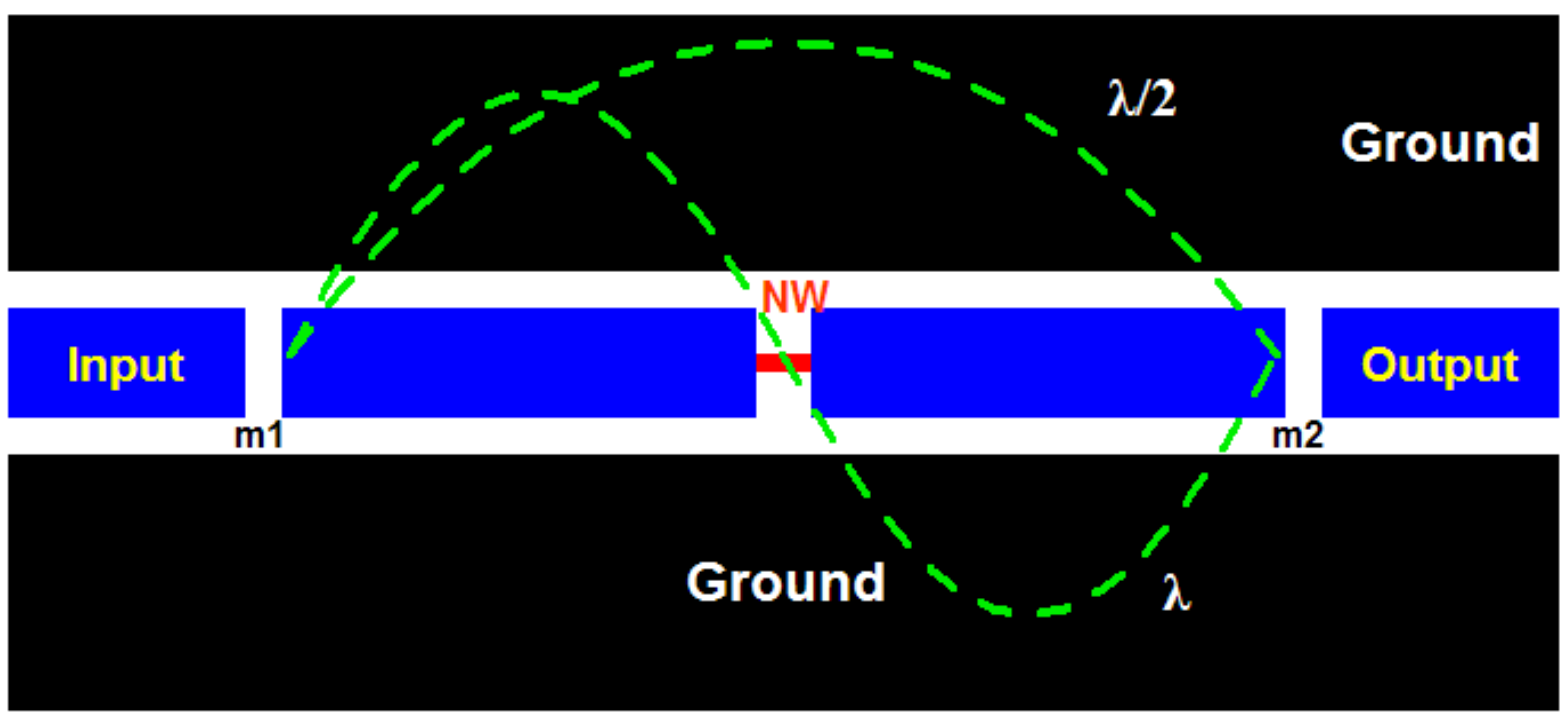

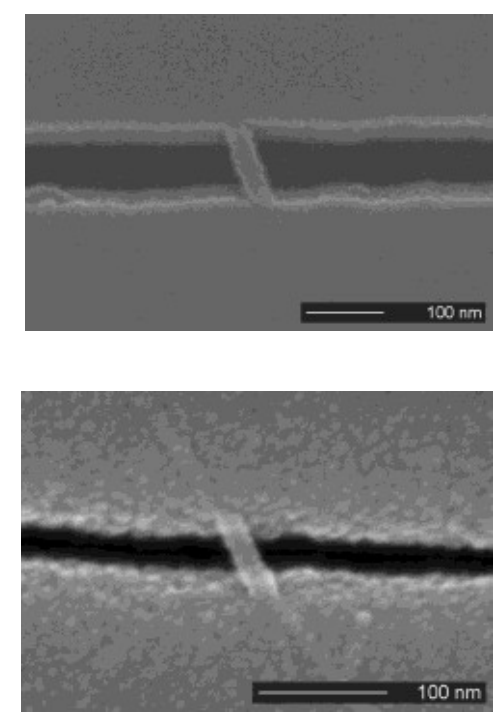

*Figure 1. (Color online) a) Schematic drawing of the measured sample. On the schematic, a Fabry-Perot resonator rendered nonlinear by inserting a thin superconducting wire at the antinode. The center conductor of the resonator (blue) is capacitively coupled to the "input" electrode, to which the microwave signal is applied. It is also coupled to the output electrode, which is used to measure the amplitude of the oscillating field. The nonlinear element, the nanowire (NW, red), is characterized by a kinetic inductance that depends on the supercurrent. The resonator is patterned from a 25 nm thick superconducting film of MoGe and the wire is produced by molecular templating [12]. The black color as well as the blue color both represent the MoGe film, but the regions shown in black are grounded, while blue regions are not. b) Scanning electron microscope images of the suspended nanowires in S1 (top) and S2 (bottom). The scale bar is 100 nm.*

## FABRICATION AND MEASURMENTS

The devices under investigation are based on a superconducting coplanar waveguide (CPW) resonator, which includes the central conductor (blue) and the ground planes (black) (Fig.1). For S1 sample, following Boaknin et al. [23], they were patterned by optical lithography in superconducting 25 nm thick MoGe film deposited in a DC magnetron sputtering system (ATC 2000 from AJA International, Inc., and a single compressed $Mo_{0.76}Ge_{0.24}$ target purchased from Super Conductor Materials, Inc.). Thus all the regions colored in blue, black, and red represent the MoGe film, while the white space in Fig.1 represents SiN deposited on an oxidized Si wafer [12]. The width of the center conductor is 20 μm and the gap between the center conductor and ground plane is 10 μm. A Fabry-Perot resonator is formed as two semi-transparent mirrors (marked "m1" and "m2" in Fig.1), about 45 fF each, are introduced into the central conductor of the CPW. These mirrors are simply a few micron wide gaps in the central conductor. The mirrors act as electric capacitors imposing a rigid boundary condition, meaning that the supercurrent through these gaps is exactly zero [24]. The length of the center conductor between the two coupling gaps is 10mm and the expected resonant frequency is ~6 GHz. The fundamental frequency (i.e. the $\lambda/2$ mode frequency) of the resonator was ~3.5 GHz and the loaded quality factor Q~500. The measured resonant frequency is much lower than expected due to the kinetic inductance contribution from the thin MoGe film. This sample is overcoupled and thus the loaded quality factor is dominated by the external dissipation due to the energy leakage through the capacitive coupling to the environment (i.e. capacitors m1 and m2). The nanowire was produced by molecular templating technique, bridging two center conductors seamlessly. The width and length of the wire is about 30 nm and 100 nm as can be seen in SEM pictures in Fig.1b, while the nominal thickness was 25 nm.

For the S2 sample, we used two layers of MoGe (a thick one first and then a thin one, after the nanotubes supporting the wire were deposited). The reason to use such double-layer technique is to create a high-Q (i.e. thick) resonator and a thin nanowire on top of it. To fabricate the sample, about 80 nm-thick MoGe film was first deposited on the SiN/SiO2/Si substrate and similarly, the molecular templating technique [12] was applied to a 10 nm MoGe film, creating a nanowire about 20 nm wide and 60 nm long. The nominal thickness of the wire was 10 nm. As for the resonator, the width of the center conductor was 10 μm and the gap was 5 μm. The coupling gaps (m1 and m2) were 4 μm in size, giving an estimated capacitance of about 1fF. The fundamental frequency of the resonator was ~4 GHz and the loaded quality factor Q~5000. Unlike the S1, this sample is undercoupled and thus the loaded quality factor is dominated by the internal dissipations.

The resonator is excited by applying a microwave signal to the input, which is coupled to the resonator through the coupling capacitor "m1" (Fig.1a). The strength of the oscillations is detected by measuring the power of the transmitted waves at the output, which escape from the resonator through the coupling capacitor m2. The desired anharmonicity of the resonator is achieved by placing a superconducting nanowire (shown as a red line in Fig.1a) at the supercurrent anti-node, i.e. in the middle of the resonator. Since the kinetic inductance should depend on the amplitude of the supercurrent, the resonance frequency is expected to be a function of the number of photons present in the resonator and therefore on the power of the driving microwave signal. The dashed green curves in Fig.1a schematically show the supercurrent amplitudes, associated with the fundamental and the first harmonic mode of the resonator, referred to as "$\lambda/2$" and "$\lambda$", respectively. These notations reflect the fact that the first resonance peak occurs when the wavelength of the plasma wave is such that $\lambda/2 = L$, and the second peak takes place at $\lambda = L$, where $L$ is the length of the resonator, i.e. the distance between the capacitors m1 and m2 in Fig.1a. In the first resonance ($\lambda/2$), the current is the maximum at the position of the nanowire, while the current through the nanowire is zero at the second resonance. Thus, one expects that the nanowire should not affect the second transmission resonant peak, but only the first one [[25],[26],[27],[28]].

The resonator measurement is based on ultra-low noise microwave techniques which have been successfully used for the readout of superconducting qubits [[29]]. In particular, these techniques allow one to control very precisely the environmental impedance and the noise seen by the nano-object under study. The input is powered by the source of a vector network analyzer (Agilent 8722D). The driving power is delivered through a stainless-steel semi-rigid cable, reduced by a 20 dB attenuator (XMA) thermalized at 800 mK and then another 10 dB attenuator held at the base temperature of 30 mK, at which the sample was held. The resonator output is immediately connected to a circulator at 30 mK, then to another circulator maintained at 800 mK. The second circulator leads to a high-electron-mobility transistor (HEMT) cryogenic ultra-low noise amplifier. This amplifier has a noise temperature of less than 4.2 K. The output of the amplifier leads to another room-temperature amplifier (MITEQ) at the top of the cryostat. There the output signal is measured by the same vector network analyzer. The network analyzer is able to sweep the power in steps of 0.1 dBm, thus allowing a detailed investigation of the dependence of the kinetic inductance of the wire on the supercurrent amplitude, as is explained in what follows.

The sample S2 was measured in He3 cryostat system, where similar microwave experimental set-up was used. At the input microwave line, 20, 3 and 12 dB attenuators (Inmet) are mounted at each temperature stage of 4, 1 and 0.3 K, respectively. At the output microwave line, two isolators are thermally anchored to 1 K and 0.3 K temperature stages for each. Then a cryogenic low noise amplifier (Low Noise Factory) is held at 4 K and two room-temperature amplifiers follow. To perform the microwave transmission measurement of the resonators, a vector network analyzer (Agilent PNA5230A) was used.

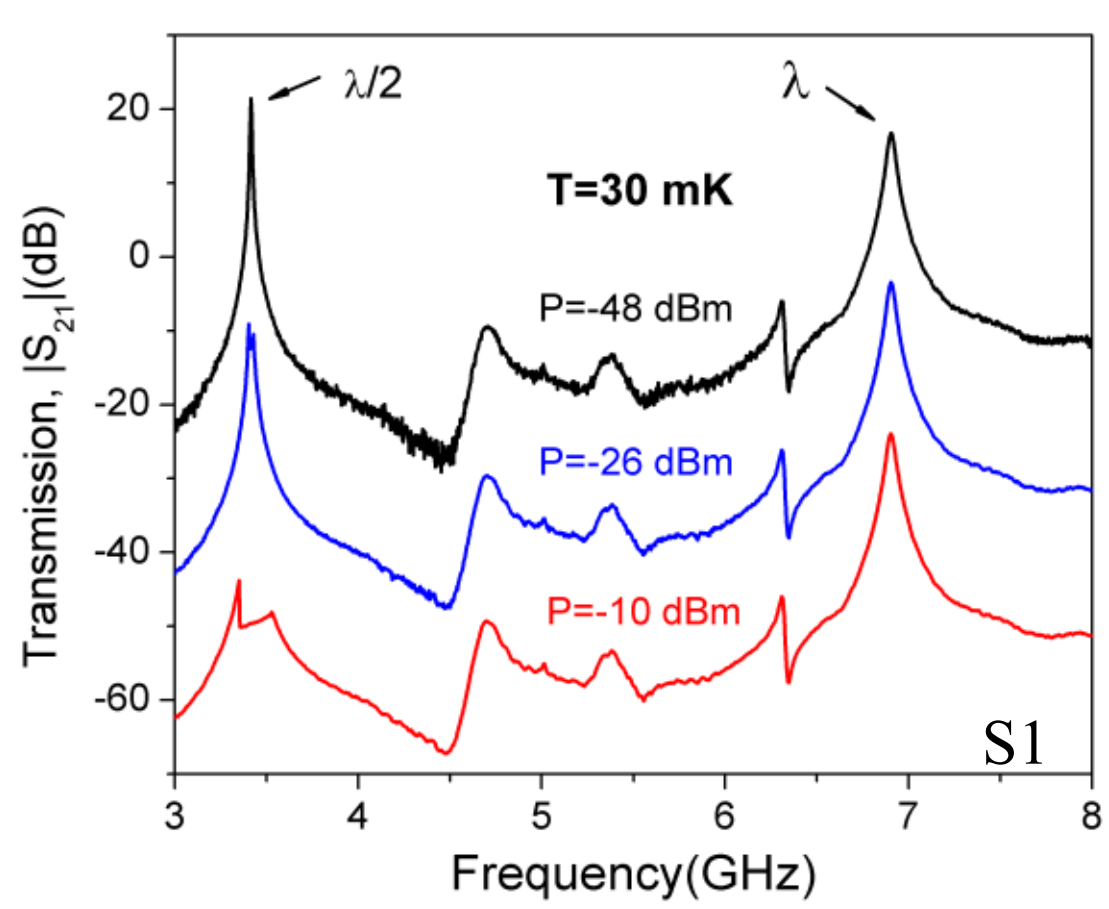

*Figure 2. (Color online)(sample S1) Transmission $S_{21}$ of the superconducting Fabry-Perot resonator with a nanowire placed in the middle of the resonator, as in Fig.1. The graph illustrates the sensitivity of the first resonance peak to the driving power* $P(\equiv P_{out}^{NA})$ *, while the second resonance is unaffected. The blue curve is shifted downwards by 20 dB and the red one by 40 dB for clarity.*

**RESULTS**

A general view of the measured transmission function of a SFPR with a nanowire is shown in Fig.2. Consider first the black curve, corresponding to the lowest driving power. Two prominent peaks are visible: the fundamental resonance at ~3.5 GHz, marked "$\lambda/2$", and the first harmonic of the resonator at ~7 GHz, marked "$\lambda$". The blue curve corresponds to a larger driving power and it shows a noticeable reduction of the transmission in the first peak, while the second peak remains unchanged. Finally, the red curve corresponds to even higher power, at which the first peak is strongly suppressed, while the second is unchanged. So it appears that at high driving power and correspondingly at a large supercurrent oscillation amplitude the first peak ("$\lambda/2$" resonance) acquires a volcano shape (i.e. exhibits a "crater"), while the second, "$\lambda$" peak, remains unchanged. To explain this one needs to compare the supercurrent patterns corresponding to the "$\lambda/2$" and "$\lambda$" resonance modes. In the $\lambda/2$ mode the supercurrent has the antinode in the middle of the resonator, exactly at the spot where the wire is located. Thus, if the $\lambda/2$ mode is excited, the oscillating supercurrent is forced to flow through the nanowire. As the amplitude is increased above the critical current of the wire, a dissipative process occurs and the resonance peak gets modified and eventually suppressed, as is evident from Fig.2. On the contrary, the $\lambda$ mode has a current node (zero) in the middle of the resonator. Thus the current through the wire is zero for this mode and the corresponding resonance peak shows no sensitivity to the activation amplitude, unless the amplitude is so large that the condensation amplitude becomes suppressed even in the superconducting film forming the main body of the resonator.

Thus in order to probe the current dependence of the kinetic inductance of the wire, we need to focus on the first resonance peak. The transformations occurring in the first peak as the power is increased are shown in detail in Fig.3a. Here the curve 1 corresponds to a low driving power (-48 dBm). At this low power a significant increase in the driving power by 4 dBm (up to -44 dBm) does not change the curve significantly (compare curves 1 and 2). As the power is increased further, the peak shifts (curve 3) to lower frequencies, and the shape of the resonance peak changes (curve 4). At the highest power, before the crater appears, the observed shift is about 3 MHz as is clear from the comparison of the curves 1 and 5. The shift is due to the fact that the kinetic inductance of the wire becomes more and more current-dependent, as the supercurrent increases. The width of the peak itself is ~7 MHz. Thus, the shift is of the same order of magnitude as the peak width. The peak width can be further reduced, if necessary, by increasing the quality factor of the resonator.

Here we emphasize that these resonance peak shifts were observed only in the fundamental mode ( $\lambda/2$ ), not in the first harmonic mode ( $\lambda$ ), clearly showing that the nonlinear effect observed in $\lambda/2$ mode is purely due to the nanowire.

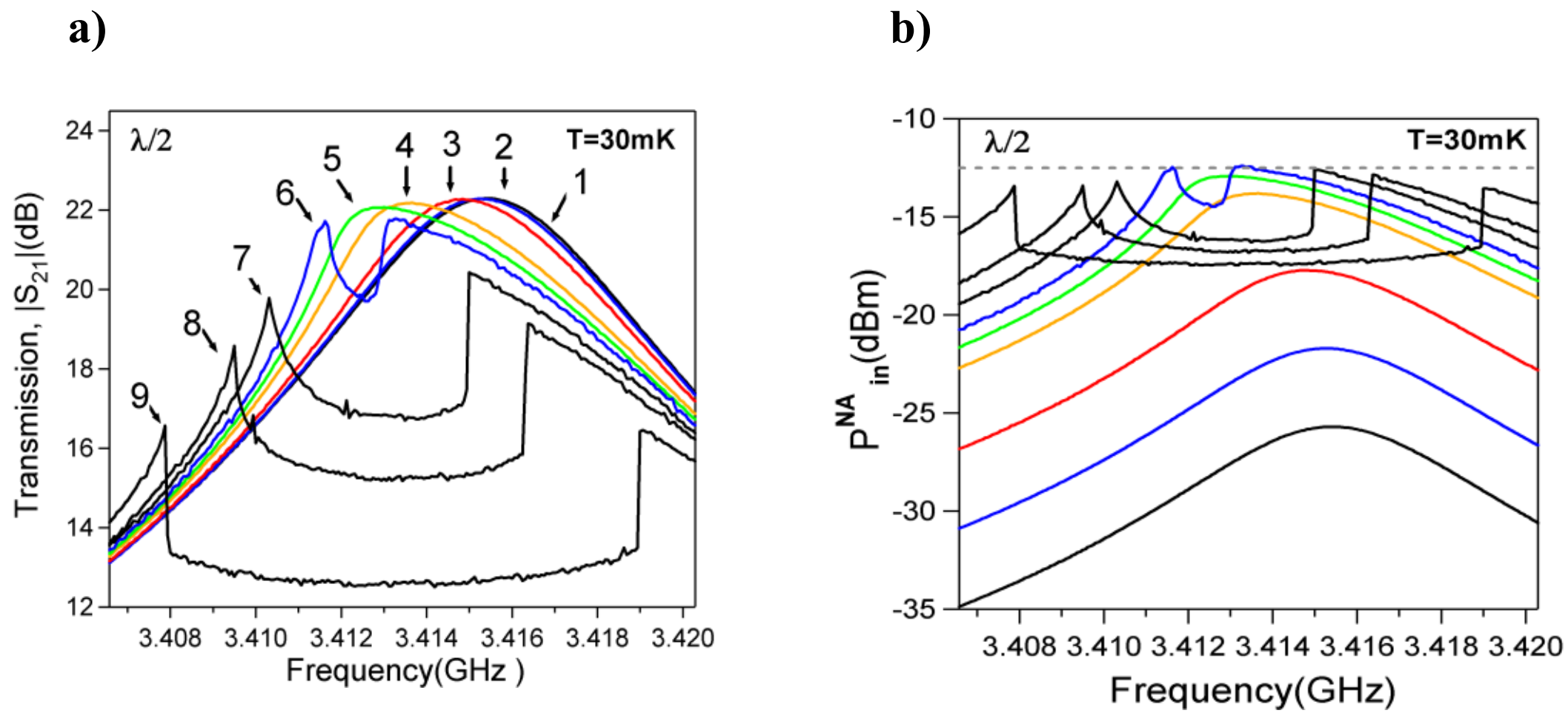

*Figure 3. (Color online) a) (sample S1) Transmission amplitude $S_{21}$ as a function of frequency for different values of the driving power. This resonance peak represents the fundamental mode of the resonator with a nanowire. The curves correspond to different driving powers, as follows: 1: $P_{out}^{NA}$ = -48 dBm (black); 2: -44 dBm (blue); 3:-40 dBm (red) 4: -36 dBm (orange); 5: -35 dBm (green); 6: -34.2 dBm (blue); 7: -33 dBm (black); 8: -32 dBm (black); 9: -30 dBm (black). b) (sample S1) Replotting of the data from (a) as the resonator output power, measured at the network analyzer input $P_{in}^{NA}$ versus frequency. Note that the square root of the output power is proportional to the amplitude of the supercurrent in the resonator. The dashed line indicates the driving power of -12.5 dBm level.*

As the power is increased further, a crater appears on top of the resonance peak, as is clear in curve 6 (and the curves with larger numbers) of Fig.3a. We suggest that the observed reduction of the transmission at the peak is due to the fact that the supercurrent amplitude inside the resonator exceeds the nanowire's critical current. As it happens, the nanowire goes to the normal state, introducing strong losses to the resonator. In what follows we will use the following notation: $P_{in}^{NA}$ will denote the power at the input of the network analyzer. $P_{out}^{NA}$ denotes the power at the output of the network analyzer, also called the driving power. This power goes, through the set of cables and attenuators, to the input of the resonator. The power transmitted through the entire circuit including the attenuators, the resonator, the circulators and the amplifiers, $P_{in}^{NA}$ is proportional to the power at the output of the resonator, which in its turn, is proportional to the square of the current amplitude in the resonator. $P_{in}^{NA}$ is plotted in Fig.3b. To plot it we use the relation $P_{in}^{NA}[dBm]=|S_{21}|[dB]+P_{out}^{NA}[dBm]$. As the driving power is increased, the overall transmitted power curves move up until the maximum reaches near -12.5 dBm line. Fig.3b shows that the current amplitude at the peak can not exceed this level, most probably the level corresponding to the critical current of the wire. This level appears slightly dependent on the value of the driving power, probably due to the fact that in this experiment the voltage, not the current, is measured, and the relation of the voltage oscillation amplitude to the current oscillation amplitude may deviated from being exactly proportional since the inductance of the wire shows some dependence on the supercurrent flowing through it. Another possible explanation might be that at higher driving powers the critical current is reached faster, thus making the occurrence of the normal state in the wire more frequent. We assume that the appearance of the crater indicates the occurrence of a significant Joule heating of the wire, as the current exceeds the critical depairing current. This, in turn, might increase the average temperature of the NW leading to the observed small deviations from the maximum peak level of -12.5 dBm. Further increase of the power leads to an increase of the crater size (curves 7, 8, and 9) and the corresponding increase of average Joule power dissipated in the wire. At low-enough driving power the crater shape appears continuous and smooth. As the power is increased, then initially the right side of the crater and then also the left side develops a discontinuity jump and a hysteresis, associated with the jump. The exact position of the jump fluctuates slightly from one sweep to the next one.

**a)** **b)**

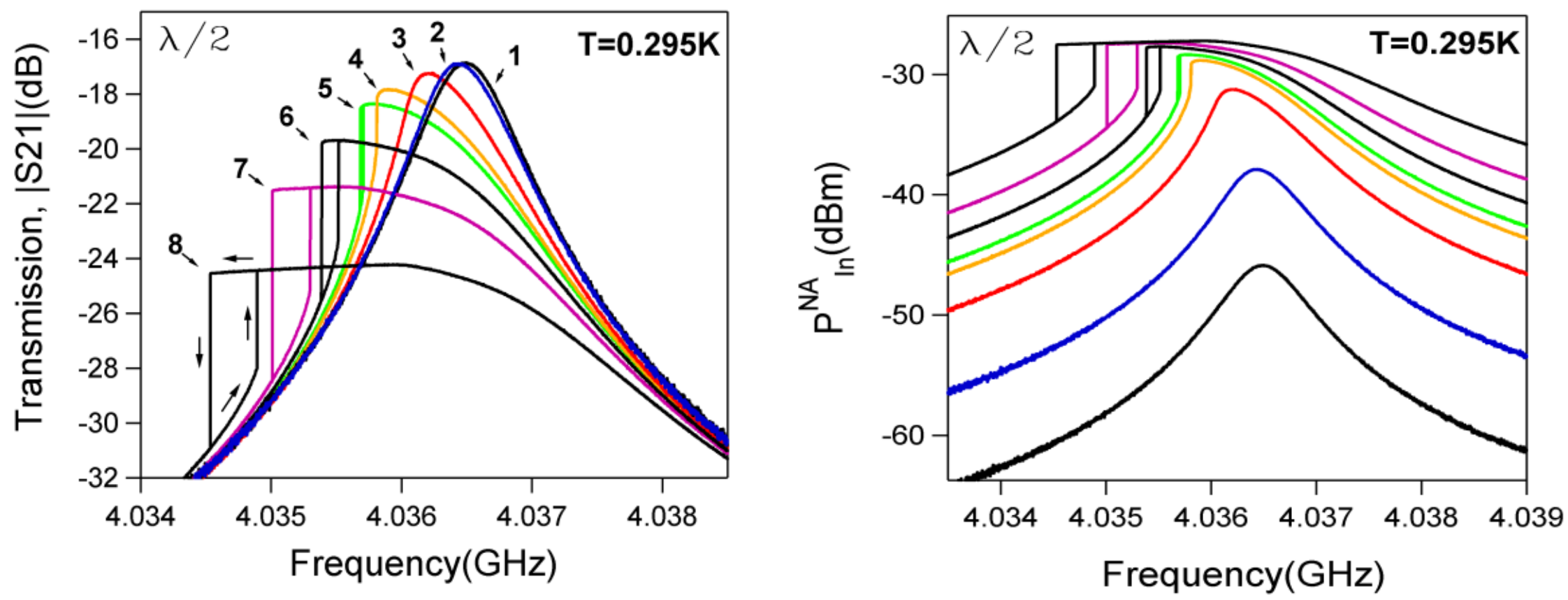

*Figure 4. (Color online) a) (sample S2) Transmission amplitude $S_{21}$(dB) in forward and backward frequency sweep for various driving powers. The graph shows Duffing bifurcation occurring at higher driving powers. The curves correspond to different driving powers: 1: $P_{out}^{NA}$ = -29 dBm (black); 2: -21 dBm (blue); 3:-14 dBm (red) 4: -11 dBm (orange); 5: -10 dBm (green); 6: -8 dBm (black); 7: -6 dBm (violet); 8: -3 dBm (black). b) (sample S2) Replotting of the data from (a) as the transmitted power $P_{in}^{NA}$ measured at the network analyzer input versus frequency.*

We have also succeeded in making a sample with a ten times larger quality factor, i.e. sample S2. Now we examine the first resonance peak of the S2 as shown in Fig.4a. The transmission amplitude was measured in both forward and backward sweep of the driving frequency. At low driving power (curve 1), the resonator response was lorentzian centered at the resonance frequency of 4.036GHz, and the quality factor was 5025. As the driving power increases, the resonance peak becomes asymmetric due to the nonlinear inductance of the nanowire (curve 1 to 4). The resonant peak shifts by ~0.5MHz (which is also of the order of the peak width). As the driving power reaches a critical power $P_{out}^{NA} = P_c$ (curve 4), an abrupt transition in the transmission amplitude, "bifurcation", was observed in both frequency sweep directions. In addition, this transition is hysteretic. As shown in curve 8 of Fig.4a, the transition occurs at higher frequency in forward sweep than in backward sweep. (The arrows indicate the direction of sweep.) These transitive and hysteretic behaviors are well-know features in a nonlinear system. The amplitude dependence of the resonance frequency in nonlinear system leads to the development of the hysteretic abrupt transitions in the Duffing model. As the amplitude increases, the nonlinear system becomes bistable at a certain frequency range, and thus the response of the cavity bifurcates. However, the downward transition in backward sweep does not occur at the same frequency where the upward transition appears since the oscillation amplitudes are different when the transitions occur. The resonance frequencies differ at the upper and lower transmission amplitude.

As in Fig3.b, Fig4.b shows the transmitted power versus frequency at different driving powers. As the driving power increases, the top part of the curves becomes flattened and also approach a certain limit of the output power, about -27.5dBm. The shape of those curves (near the limit output power) is quite different from that of the Duffing model as will be discussed below. This behavior indicates the maximum supercurrent amplitude inside the cavity continues to increase as the driving power become strong but to converge to a certain value, which is probably the critical depairing current. The qualitative difference between S1 and S2 is that S1 is strongly overcoupled and has a relatively low quality factor, while S2 is undercoupled and has an about ten times higher quality factor.

## MODEL

The results in Fig.3 and Fig.4 make it clear that the shape of the resonance peak depends on the driving power considerably. In order to examine the nonlinear aspect of the nanowire, we consider only the transmitted output power $P_{in}^{NA}$ vs. frequency for the driving powers that do not create the craters, but give asymmetric peaks as shown in Fig.5a. The driving power ranges from -48 dBm to -34.4 dBm, where the latter was the highest driving power that does not create the crater. (The power of -34.3 dBm was sufficient to produce a crater.) To understand our nanowire-resonator system, we model it as a Duffing nonlinear system with a cubic nonlinearity [20]. It is well-known that in the weak nonlinear limit, the frequency-response curves, i.e. stationary solutions, can be obtained analytically by solving the nonlinear equation of motion approximately. We can describe the nanowire-resonator as an equivalent lumped series effective LRC circuit near the resonant peak, and solve the equation of motion as in [[30]] to reproduce the transmitted output power versus frequency curves with the driving power given by the settings of the network analyzer. In this approximate model (see Appendix for details), a nanowire is considered a nonlinear non-dissipative inductive element, and is assumed to hold a sinusoidal current-phase relationship (CPR) $I(\delta) = I_0 \sin(\delta)$. This is justified by the fact that in the approximate solution given in [30] only the linear and the cubic terms of the CPR are retained anyway, and for a thin superconducting wire one expects that the CPR holds only a linear and a cubic term [18]. The model has five adjustable fitting parameters: $I_0$ (critical current), $\omega_0$ (resonant frequency), $Q_L$ (loaded quality factor, the same one for all values of the driving power), and $K_1$ and $K_2$ (scaling factors representing the attenuation of attenuators and the semi-rigid coaxial cables connected to the input of the resonator, and the combined effect of the cables, circulators and amplifiers connected to the output of the resonator, respectively). With the model we could fit the whole family of the output power curves as a function of frequency at each driving power. The equations used for making the fitting are given in the Appendix below. The fitted curves are in a quantitative agreement with the data as shown in Fig.4a. In the fitting procedure, we note that the three parameters ($I_0$, $K_1$, and $K_2$) are not determined uniquely. One of them can be arbitrarily chosen to find two others. The best fits could be achieved with any choice of $K_1$ if other fitting parameters are adjusted appropriately. From the theoretical fitted curves, we can obtain directly the current amplitudes at each given frequency and drive power. In the experiment we see that at some critical driving power (which actually was -34.4 dBm) a crater develops. We denote the corresponding supercurrent amplitude as $I_{C.M}$. Our interpretation is that this is the "measured" critical current. Thus we use the condition $I_0 = I_{CM}$. The best fit, for the entire set of curves in Fig.5a, was achieved with $I_0 = I_{CM} = 4.1\mu A$, $Q_L = 515$, $\omega_0/2\pi = 3.4156\,GHz$, $K_1 = 2.37\times10^{-5}$, and $K_2 = 44400$. The $\omega_0$ and $Q_L$ are consistent, within less than 1% deviation, with those of the curve corresponding to the lowest driving power, where the nonlinear effect is negligible. Thus it is concluded that a nonlinear regime does exist, in which the superfluid density is suppressed in the wire, but the dissipation is not increased. In other words, the nanowire can act as dissipationless nonlinear inductor. The fitting parameter $K_1 = 2.37 \times 10^{-5}$ corresponds to 46.3dB attenuation on the circuit connected to the input of the resonator, which is close to the expected value based on our approximate knowledge of the cable attenuations and the attenuators placed in the resonator input line (~40dB) . The value of the wire's critical current $I_0$ is roughly consistent with the expectations based on the nominal sputtered film thickness. However, the value $I_0$ can not be determined exactly from the known thickness of the sputtered MoGe since the critical current depends on the actual size of the nanotube or a nanotube rope templating the wire [21]. Since the size of the tubes/ropes is not fixed and it is known only very roughly, we can not make a precise prediction of the wires critical current independently of the data and the fitting procedure of Fig.5a.

**a)** **b)**

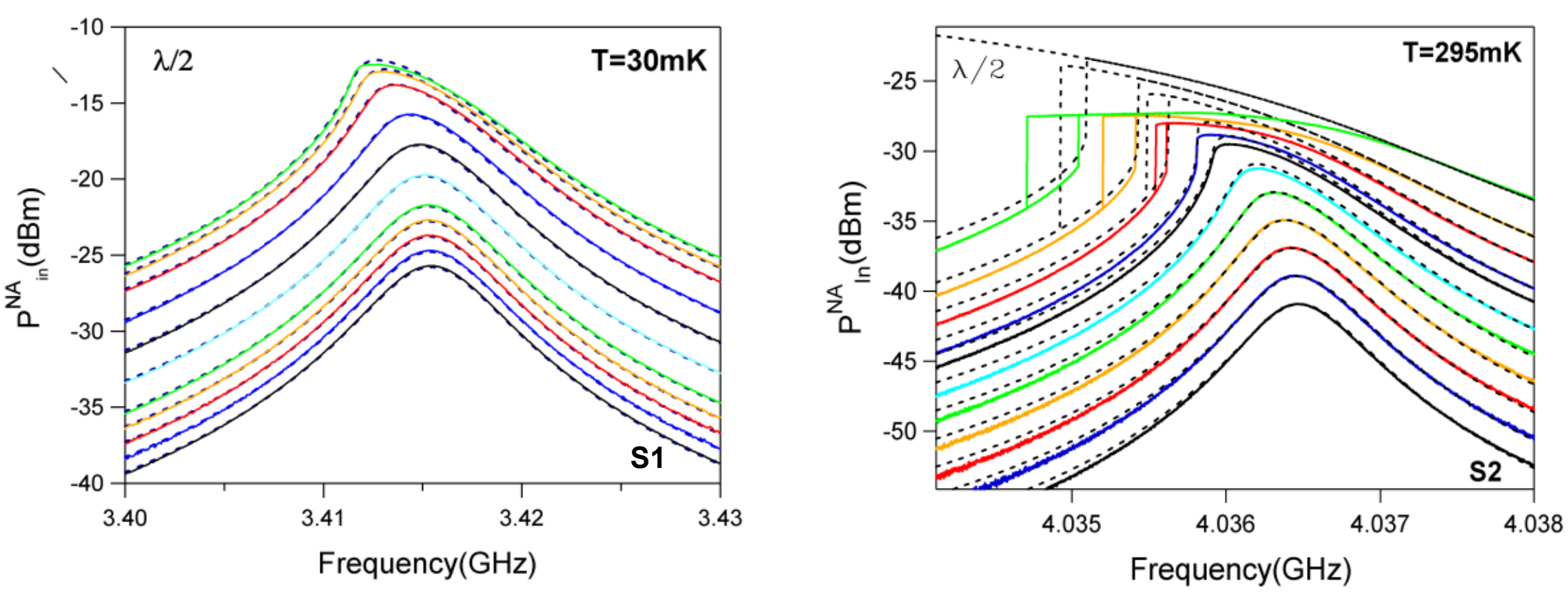

*Figure 5. (Color online) Transmitted power $P_{in}^{NA}$ versus frequency is plotted for samples for S1 (left) and S2 (right), for the corresponding fundamental modes. The parameter for the presented family of curves is the nominal driving power. The solid curves represent data, and the dashed lines are fits to the data. a) The driving power* ranges *from -48dBm to -34.4dBm, starting from the bottom curve. The driving powers used are: $P_{out}^{NA}$ = -48, -47, -46, -45, -44, -42, -40, -38, -36, -35, -34.5, and -34.4 dBm. b) The driving powers used are $P_{out}^{NA}$ =-24, -22, -20, -18, -16, -14, -12, -11, -9, -7, and -4 dBm*

We observe that the agreement between the model given by Eq.(3) and the data is very good, except for the two topmost curves corresponding to the highest excitation powers, and the deviations are observed only very close to the resonance maximum. The deviations might be due to the fact that the wire CPR deviates from the sinusoidal one as the current approaches the critical current.

Here we should stress the fact that the fitting procedure, which provides good agreement with the date, uses a nonlinear model with the quality factor independent of the driving power. Otherwise, there would be a broadening of the resonance peak and the suppression of the maximum peak height. If Bogoliubov quasiparticles [31] and/or Little phase slips [32] were creating a significant dissipation, we would observe a dependence of the quality factor on the driving power. Yet the quality factor appears independent of the driving power. Thus a low level of dissipation in the nanowire is strongly suggested by these measurements and the modeling results, even in the regime when the kinetic inductance changes significantly due to strong supercurrent amplitude. This non-dissipative nature is a key requirement for implementing a qubit. Therefore, the results provide evidence that the nanowire as a nonlinear and non-dissipative element can be used for making qubits.

For the S2, we performed a similar analysis on the data as shown in Fig.5b. At the low and intermediate driving powers, at which the bifurcation does not occur, the fits are in reasonable agreement with the data. However, beyond the critical power where the bifurcation starts to develop, the fits deviate from the data. The first noticeable discrepancy is the shape of the curves: The top part of the experimental curves become flattened out. As the driving power is increased, the flattened part raises only slightly. The second difference is the size of the hysteresis of the bifurcation. The size of the hysteresis of the data is much smaller than that of the fit. This correspond to he fact that the jump from the high transmission branch to the low transmission branch, as the frequency is scanned downwards, happens earlier than the model predicts. We suspect that the observed deviations are due to the fact that the depairing current of the wire is reached at powers slightly higher than the power at which the dissipation occurs.

Those two differences are not well explained by the Duffing model. In order to investigate this discrepancy, we performed numerical analysis of the lumped series effective LRC circuit with the nanowire. The nonlinear second-order differential equation was solved for $\varphi$ (phase across the nanowire) iteratively in time domain using the fourth order Runge-Kutta method and then the transmission output power of the cavity vs frequency curves were obtained (see Appendix B). To represent the dynamics of the bifurcation in the system more effectively, we focus on the frequency locations $\omega_D$ where the maximum derivative and the maximum amplitude occur in each frequency-response curve. Now we can plot the frequency-response data (Fig.5b) in the plane of normalized driving power $\left(P/P_c\right)$ and normalized frequency $\left(\Omega\right)$, where $P$ :driving power, $P_c$ :critical power where the bifurcation of transmission amplitude appears, $\Omega = \frac{\delta\omega}{\Gamma} = \frac{\left(\omega_0 - \omega_D\right)}{\omega_0/2Q_L}$ , $\omega_0$ : resonant frequency, $\omega_D$ :frequencies at the maximum(minimum) derivative of the transmission curve versus the driving frequency and/or the frequency of the maximum of the transmission amplitude, and $Q_L$ is the loaded quality factor. Then the so-called bistability diagram is drawn in the plane of $\left(\Omega, P/P_c\right)$. The region where $P/P_c(dB) > 0$ is called bifurcation region, and the region where $P/P_c(dB) < 0$ is called sub-bifurcation region.

Now we first show the bistability diagram of the Duffing oscillator. From the stationary solution of the Duffing model, the upper/lower bifurcation branches in the bifurcation region, the maximum derivative and the maximum amplitude branches in sub-bifurcation region are given in [23]. This bistability diagram of the Duffing system is plotted in Fig.6a as solid lines. There are two branches, upper(blue) and lower(red) branches in the bifurcation region. The upper(lower) branch corresponds to the jump-up(jump-down) of the transmission amplitude in the forward(backward) frequency sweep. In the sub-bifurcation region, the lower(red) curve corresponds to the maximum derivative of the response curve, and the upper(black) curve corresponds to the maximum amplitude. Those are universal such that when the transmission amplitude vs frequency data are properly rescaled, they all fall on the universal curves. The measurement data is also plotted in Fig.6a as discrete symbols. The data fall on the theoretical curves very well, except for the lower bifurcation branch (which represents the jump from the high oscillation amplitude of the nonlinear oscillator to the lower amplitude and which could be influenced by the close proximity of the depairing current of the wire). The lower bifurcation branch of the data lies above the theoretical lower bifurcation branch, indicating the size of hysteresis is smaller compared with that of the Duffing system. This discrepancy in lower bifurcation points can be explained, for example, by power-dependent dissipation at high current, approaching the depairing current. This is because the downward transition (corresponding to the lower bifurcation branch) occurs from a higher supercurrent to a lower supercurrent. This regime of very high currents should be avoided if the wire is to be used in a qubit setting.

Now for our numerical results, we first numerically solved the Duffing equation to obtain transmission output power vs frequency plots (Eq.3 in Appendix A). The bistability diagram of those numerical solutions exactly matched that of analytic solutions of the Duffing model. Next, we numerically solved the LRC circuit with a Josephson junction without taking only the lowest order of nonlinearity. The diagram of the Josephson junction case is almost identical to the theoretical one, but the lower branch slightly deviates, which is not surprising since the theoretical diagrams are derived from the Duffing system that takes approximation up to the first order of $(I_s/I_0)^2$. Next, we replaced the Josephson junction with the nanowire as a nonlinear inductance, where the current phase relationship of a long nanowire $I_s = 3\sqrt{3}I_0/2\left[\varphi/(L/\xi) - \left(\varphi/(L/\xi)\right)^3\right]$ is used[18].( $I_0$ :critical current, $\varphi$: phase across the wire, $L$: length of the wire, and $\xi$ :coherent length. We used $I_0 = 2\mu A$ and $L/\xi = 7$ .) We tested different degree of nonlinearity of the nanowire and, however, all of them were close to the theoretical Duffing bistability curves, although some deviations we observed in the lower branch of the data. Then we added power-dependent dissipation term $R = R_0\left(1+\alpha(I_s/I_0)^2\right)$ to the model, where $R_0 = 0.0099$ , $I_0 = 2\mu A$, $I_s$ is the time-dependent supercurrent through the nanowire. We should note that, in our model, the Fourier component of the supercurrent through the wire only at the driving frequency is extracted and the square of that component is assumed to be proportional to the transmitted power of the cavity. The best fit was obtained with $\alpha = 60$ showing a reasonable agreement with the data. The final simulation results are plotted in Fig.6b as solid lines. This result indicates that at very high ac current amplitude the internal dissipation in the nanowire is not negligible. It is assumed that the dissipation mainly takes place in the nanowire rather than in the resonator since the current density is much higher in the nanowire as the width of the nanowire is about 200 times smaller than that of the center conductor of the resonator. Lastly, we should mention that the simulated transmission output power vs frequency curves (not shown) were not able to reproduce the observed flattening of the peak of the transmission curve at high driving powers (Fig.5b).

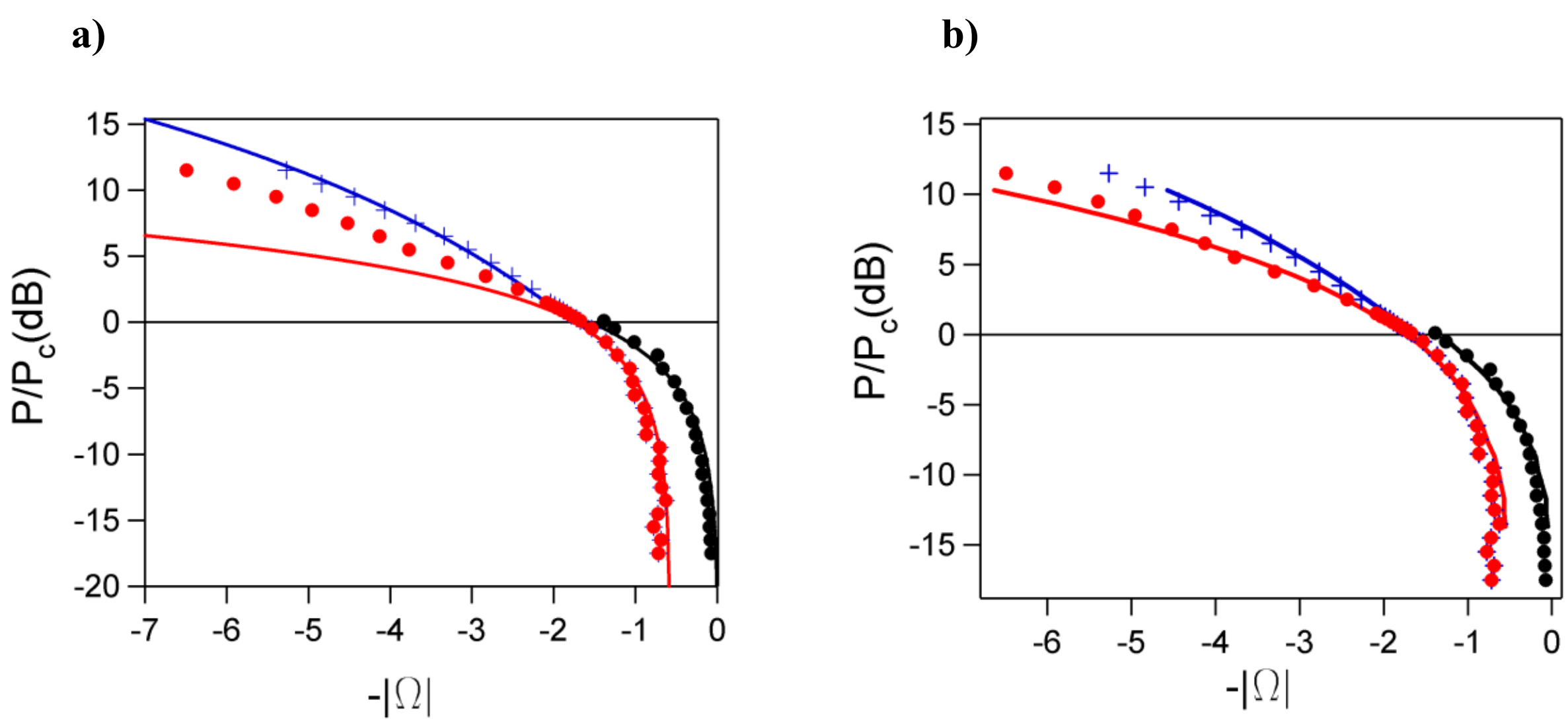

*Figure 6.(Color online) Bistability diagrams of the Duffing model, the data for S2, and the simulation results. The curves in blue/red and black correspond to maximum derivative and maximum amplitude in frequency-response curve, respectively. a) Bistability diagrams of the approximate analytic solution (solid lines) of the Duffing model explained in Appendix A [19] and the data (crosses and dots) are depicted. b) Bistability diagrams of our numerical simulation results explained in Appendix B (solid lines) and the data (crosses and dots) are shown.*

## QUBIT DESIGN PROPOSAL

We establish that by placing a nanowire into a FP resonator it is possible to make the system nonlinear, while the dissipation remains unchanged in a certain range of supercurrent amplitudes. In the future work, such resonators with inserted nanowires will be tested for the pursuit of a qubit, namely the proposed nanowire-FP qubit. To our knowledge, this new qubit type has not been demonstrated yet. For this nanowire-resonator system to work as a qubit, it should be unharmonic enough to satisfy the condition $\delta\omega/N > \Gamma$, where $\delta\omega, N$ and $\Gamma$ are the maximum resonance shift in the $\lambda/2$ mode, the number of photons in the cavity corresponding to the shift, and the bandwidth of the resonance peak at -3 dB of the maximum of the transmission. In our experiments so far, we estimated that $\delta\omega$ =3MHz, $N \sim 10^4$, $\Gamma$ ~6MHz for S1, and $\delta\omega$ =0.6MHz, $N \sim 10^3$, $\Gamma$~0.8MHz for S2 and these have not met the criteria above yet [33]. Therefore, the task to increase the nonlinearity of a nanowire should be performed by reducing the thickness of the nanowire or making a constriction in the nanowire. Making constrictions is possible using a highly focused high-energy electron beam of a transmission electron microscope [34,35]. Also, he critical current can be increased by pulsing [22]. The quality factor $Q$ should be increased up to $10^5 \sim 10^6$, which has been achieved in superconducting coplanar waveguide resonators [36,37].

The schematic of the proposed qubit is presented in Fig.7. Two resonators are used. The main resonator is horizontal in the drawing. It will be used for the qubit readout. The qubit resonator is placed vertically in the drawing. It has two nanowires in the middle, i.e. at the antinode of the supercurrent of the fundamental mode of the resonator. By applying a magnetic field perpendicular to the plane of the device, it should be possible to control the frequency of the qubit resonator, due to the critical current of the wires modulated [16]. If the qubit is operational, one expects to observe a splitting of the resonance peak of the main resonator and the dependence of the splitting on the number of photons in the qubit resonator. The equivalent scheme, valid near the resonance, is shown in Fig.5b. The first step to characterize the qubit would be to observe the two well-defined quantum states using a well-established experimental technique, namely the frequency-domain approach as was used in c-QED experiments by Wallraff [38].

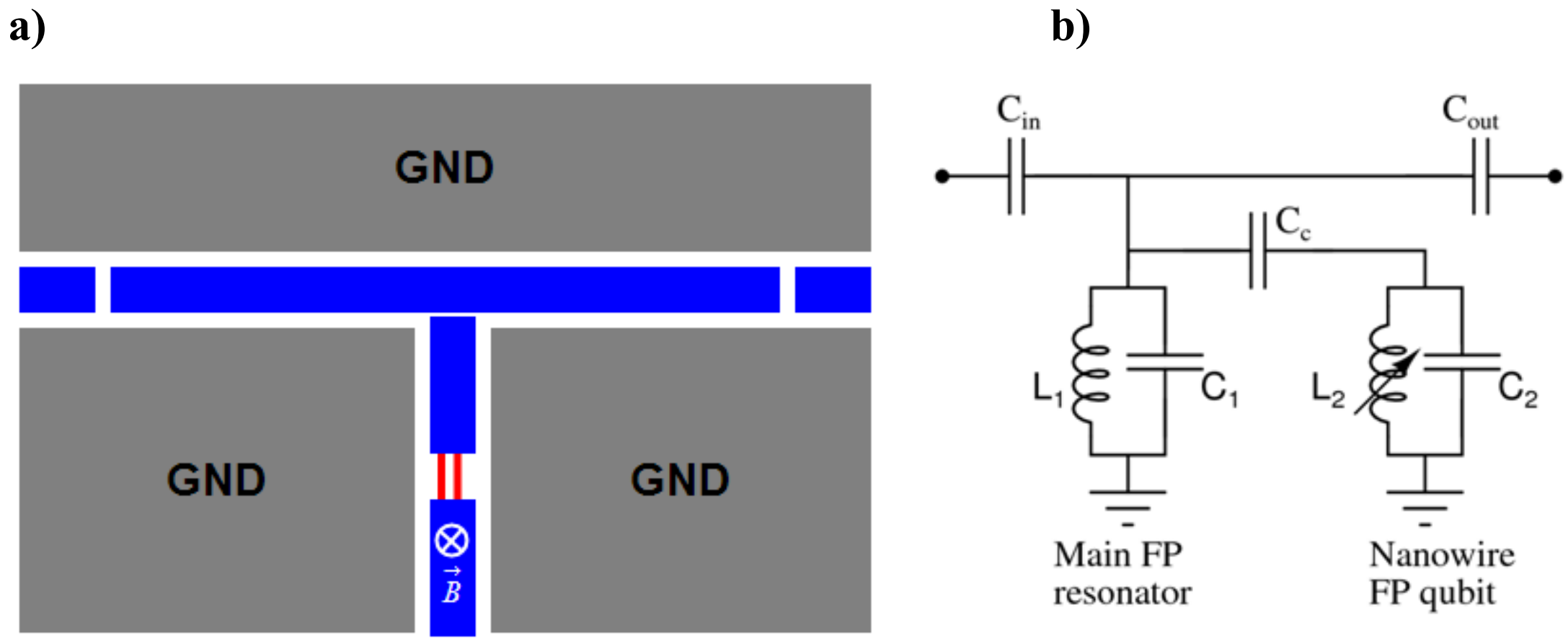

*Figure 7. (Color online) a). Schematic of a nanowire-FP qubit (vertical) coupled to the main resonator (horizontal). The qubit is a FP-type coplanar waveguide resonator having two parallel nanowires (red) in the center. The nanowires make the resonator anharmonic. Thus two levels can be addressed by a proper choice of frequencies. The critical current of the pair of nanowires can be controlled by the perpendicular magnetic field[16].Thus the qubit resonator can be tuned in resonance with the main resonator, if desired. The ground planes (gray) are indicated by "GND." Although the bottom end of the qubit resonator is shown ungrounded, it can be link to the ground plane if desired. Note that both the gray color and the blue color represent the MoGe film, but the regions shown in gray are grounded, while blue regions are not. b) Simplified equivalent circuit of the sample. The inductor in the qubit can be tuned with the magnetic field, and it is current-dependent, thus termed "nonlinear". The arrow crossing the inductor symbolizes the current dependence of the inductor.*

It should be noted that the qubit design outlined here is qualitatively different from the design proposed by Mooij and Harmans (MH), which is also based on nanowires [39,40]. The MH design relies essentially on the presence of quantum phase slips QPS in the nanowire. Yet the existence and properties of QPS are not yet firmly established and represent a subject of intensive research [[41], [42], [43], [44], [45], [46], [47], [48], [49, 50], [51]].

The qubit design proposed here does not require QPS at all. On the contrary, it relies only on the dependence of the kinetic inductance of the inserted nanowire on the value of the supercurrent oscillation and should show the best performance if QPS is absent. Thus a comparative study of the MH and our qubits can provide, among other things, definitive evidence in favor or against the existence of coherent QPS in superconducting nanowires.

In conclusion, we have demonstrated the nonlinear inductance of the nanowire in the Fabry-Perot resonator by measuring the transmission signal at a range of driving powers. The nanowire-resonator was modeled using lumped series effective LRC elements, and the transmission behavior near the fundamental resonance peak was well-explained by a Duffing oscillator model. We also proposed a qubit design for the nanowire-FP qubit.

**ACKNOWLEDGMENTS**

We would like to thank M. Devoret for useful discussions and for suggesting the idea for this project. This work was supported by the U.S. Department of Energy under Grant No. DO-FG02–07ER46453. This work was carried out in part in the Frederick Seitz Materials Research Laboratory Central Facilities, University of Illinois, which are partially supported by the U.S. Department of Energy under Grant Nos. DE-FG02–07ER46453 and DE-FG02–07ER46471.

**APPENDIX A: Equation of motion of the lumped series effective LRC circuit with a Josephson junction**

Following the thesis of Metcalfe [30], we model the resonator as a lumped series effective LRC circuit. The model is only good near the resonance peak. The model allows us to calculate oscillating charge amplitude $2|A|$ and the supercurrent amplitude $2|A|\omega$ as functions of the frequency of the driving signal $\omega$ and the voltage amplitude of the driving signal $V_d = \sqrt{R_1 P_{in}^{\mathrm{Res}}[W]}$ . The oscillator with the junction nonlinear inductor is described by the following nonlinear equation.

$$\left( L_{eff} + \frac{L_J}{\sqrt{1-\dot{q}^2/I_0^2}} \right)\ddot{q} + R_{eff}\dot{q} + \frac{q}{C_{eff}} = V_{eff}\cos(\omega t) \qquad (1)$$

,where the effective charge on the effective capacitor depends on time as $q(t) = A(t)e^{i\omega t} + \overline{A(t)e^{i\omega t}}$ . By expanding the nonlinearity term to the lowest order $1\Big/\sqrt{1-\dot{q}^2/I_0^2} \cong 1+\left(\dot{q}^2/2I_0^2\right)$, we acquire

$$\left(L_T + \frac{\dot{q}^2}{2I_0^2}\right)\ddot{q} + R_{eff}\dot{q} + \frac{q}{C_{eff}} = V_{eff}\cos(\omega t) \tag{2}$$

The frequency response curve [30] can be approximately written as

$$\left(\frac{1}{\Omega^2} + \left(|B|^2 - 1\right)^2\right)|B|^2 = \beta \tag{3}$$

, where

$B = \frac{A\omega}{I_0}\sqrt{\frac{1}{2\Omega\varepsilon^2}}$, $\beta = \frac{V_{eff}^2}{\phi_0^2\omega^2}\left(\frac{1}{2\Omega\varepsilon^2}\right)^3$, $\Omega = \frac{\delta\omega}{\Gamma}$, $\Gamma = \frac{\omega_0}{2Q_L}$, $\varepsilon = \sqrt{\frac{L_T}{L_J Q_L}}$, $L_T = L_{eff} + L_J$,

$L_{eff} = \frac{Z_0}{4\omega_0/2\pi}$ , $L_J = \phi_0/I_0$ , $\phi_0 = \hbar/2e$ , $V_{eff} = Z_0\omega_0 C_{in} V_d$ , $V_d = \sqrt{R_1 P_{in}^{\mathrm{Res}}[W]}$ ,

$P_{in}^{\mathrm{Res}}[W] = K_1 * P_{out}^{NA}[W]$ , $P_{in}^{NA}[W] = K_2 Z_0 (2\omega A)^2$ , $P_{out}^{NA}[W] = 0.001*10^{P_{out}^{NA}[dBm]/10}$ ,

$P_{out}^{NA}[dBm] = 10\log_{10}(P_{out}^{NA}[W]/0.001)$ , and $Q_L \approx Q_{ext} = \frac{\pi}{2Z_0 R_L \omega_0^2 (C_{in}^2 + C_{out}^2)}$.

**Other Notations:**

$h$ - Planck constant, $e$ - the electronic charges.

$q$ - Electric charge in the series LRC circuit.

$Z_0$ - Characteristic impedance of the coplanar waveguide used to make the FP resonator. It is estimated to be 82.5Ω.

$\omega_0/2\pi$ - Resonant frequency.

$\delta\omega = \omega - \omega_0$

$I_0$ - Critical current of a nanowire or Josephson junction.

$C_{in}$ and $C_{out}$ correspond to the input and output coupling capacitors of the resonator, i.e. the values of the capacitors “m1” and “m2” in Fig.1, respectively. In our sample, $C_{in} = C_{out} \sim 45\text{fF}$.

$V_d$ : Voltage amplitude of the microwave source driving the resonator.

$V_{eff}$ : Effective voltage amplitude of the driving source in the series LRC circuit.

$L_{eff}$ , $R_{eff}$ and $C_{eff}$ - Inductance, resistance and capacitance in the series LRC circuit effectively representing the resonator(without the nanowire), respectively.

$P_{in}^{NA}[dBm/W]$ - Transmitted output power of the resonator at the input of the network analyzer in *dBm or Watt.*

$P_{in}^{\mathrm{Re}\,s}[dBm/W]$ - Power at the resonator input in *dBm or Watt.*

$P_{out}^{NA}[dBm/W]$ - Network analyzer output power (we call it "driving power") in *dBm or Watt.*

$R_1 = 50\Omega$ is the source impedance (i.e. the impedance of the circuit connected to the input of the resonator).

$R_L = 50\Omega$ is the load impedance (i.e. the impedance of the circuit connected to the output of the resonator).

$Q_{ext}$ and $Q_L$ - External and loaded quality factor.

$K_1$ - Unitless scaling factor between $P_{in}^{\mathrm{Re}\,s}[W]$ and $P_{out}^{NA}[W]$, i.e., $P_{in}^{\mathrm{Re}\,s}[W] = K_1 * P_{out}^{NA}[W]$.

$K_2$ - Unitless scaling factor that relates the energy store in the resonator and the energy reaching the input of the network analyzer, as $K_2 = P_{in}^{NA}[W]/Z_0(2\omega A)^2$.

**Fitting parameters:** $I_0$, $\omega_0$, $Q_L$, $K_1$ and $K_2$.

## APPENDIX B: Equation of motion of the lumped series effective LRC circuit with a nanowire

The lumped series effective LRC circuit with a nanowire is described by the following equation.

$$L_{eff}\ddot{q} + R_{eff}\dot{q} + \frac{q}{C_{eff}} + V_{NW} = V_{eff}\cos(\omega t) \tag{4}$$

Here and everywhere $V_{NW}$ is the voltage between the ends of the nanowire, and $q$ is the charge on the capacitor. In order to solve for the phase difference between the ends of the nanowire, $\varphi(t)$, we take time derivative of the equation (4), and substitute for $V_{NW}$ , $\dddot{q}$ , $\ddot{q}$ ,

and $\dot{q}$, using the following relations $\dot{V}_{NW} = (\hbar/2e)\ddot{\varphi}$, $\dot{q} = I(\varphi)$, $\ddot{q} = (dI/d\varphi)\dot{\varphi}$, and $\dddot{q} = (d^2I/d\varphi^2)\dot{\varphi}^2 + (dI/d\varphi)\ddot{\varphi}$. Here and everywhere $I(\varphi)$ is the supercurrent through the nanowire. Then we obtain the ordinary second-order differential equation and solve for $\varphi(t)$ using the 4$^{\text{th}}$ order Runge-Kutta method.

$$\ddot{\varphi}\left(L_{eff}\left(\frac{dI}{d\varphi}\right)+\frac{\hbar}{2e}\right)+L_{eff}\left(\frac{d^2I}{d\varphi^2}\right)\dot{\varphi}^2+R_{eff}\left(\frac{dI}{d\varphi}\right)\dot{\varphi}+\frac{I(\varphi)}{C_{eff}}=-V_{eff}\,\omega\sin(\omega t) \qquad (5)$$

**References**

* These authors contributed equally to this work.

[24] The normal current is, of course, also zero, but it is not taken into consideration since the plasma oscillation considered here is analogous to the Mooij-Schön mode and involves only the condensate, not the quasiparticles. It should also be mentioned that quasiparticles, if present, would provide a source of dissipation and would reduce the quality factor of the resonator.

[27] C. C. Chin, D. E. Oates, G. Dresselhaus, and M. S. Dresselhaus, *Phys. Rev. B* **45**, 4788 (1992).

[28] E. A. Tholén, et al., *Appl. Phys. Lett* **98**, 253509 (2007).

[29] I. Siddiqi, R. Vijay, M. Metcalfe, E. Boaknin, L. Frunzio, R. J. Schoelkopf, and M. H. Devoret, *Phys. Rev. B* **73**, 054510/1-6 (2006).

[30] M. B. Metcalfe, Ph. D thesis, Yale University (2008).

[31] Michael Tinkham, "Introduction to superconductivity", Dover Publications, 2004

[32] W. A. Little, *Phys. Rev.* **156**, 396 (1967).

[33] The average number of photons in the cavity, $\langle N \rangle$ is estimated by $P_{out}^{\mathrm{Res}} = \langle N \rangle \hbar\omega_0 \Gamma$, where $P_{out}^{\mathrm{Res}}$ : output power from the resonator, and $\Gamma$: linewidth of the resonator given by $\Gamma = \omega_0 / Q_e$ = resonant frequency / external quality factor.

[34] C. J. Lo, T. Aref, and A. Bezryadin, *Nanotechnology*, **17**, 3264-3267 (2006).

[35] T. Aref, M. Remeika, and A. Bezryadin, *J. Appl. Phys.* **104**, 024312 (2008).

[36] Luigi Frunzio, et al., *IEEE Transactions on Applied Superconductivity*, **15**, 860 (2005).

[37] M. Göppl, et al., *J. Appl. Phys*. **104**, 113904 (2008).

[38] A. Wallraff, D.I. Schuster, A. Blais, L. Frunzio, R.-S. Huang, J. Majer, S. Kumar, S.M. Girvin, and R.J. Schoelkopf, *Nature* **431**, 162-167 (2004).

[39] J. E. Mooij and C. J. P.M. Harmans, *New Journal of Physics* **7**, 219 (2005).

[40] J. E. Mooij and Y. V. Nazaron, *Nature Physics* **2**, 169 (2006).

[41] N. Giordano, *Phys. Rev. Lett.* **61**, 2137 (1988); N. Giordano and E. R. Schuler, *Phys. Rev. Lett.* **63**, 2417 (1989); N. Giordano, *Phys. Rev. B* **41**, 6350 (1990); N. Giordano, *Phys. Rev. B* **43**, 160 (1991); N. Giordano, *Physica B* **203**, 460 (2001).

[42] A. Bezryadin, C. N. Lau, and M. Tinkham, *Nature* **404**, 971 (2000).

[43] C. N. Lau, N. Markovic, M. Bockrath, A. Bezryadin, and M. Tinkham, *Phys. Rev. Lett*. **87**, 217003 (2001).

[44] K. A. Matveev, A. I. Larkin, and L.I. Glazman, *Phys. Rev. Lett*. **89**, 096802 (2002).

[45] A. D. Zaikin, D. S. Golubev, A. van Otterlo, G. T. Zimányi, *Phys. Rev. Lett*. **78**, 1552 (1997).

[46] G. Refael, E. Demler, Y. Oreg, and D. S. Fisher, *Phys. Rev. B* **68**, 214515 (2003).

[47] H. P. Büchler, V. B. Geshkenbein, and G. Blatter, *Phys. Rev. Lett*. **92**, 067007 (2004).

[48] M. Kociak, A. Yu. Kasumov, S. Gueron, B. Reulet, I. I. Khodos, Yu. B. Gorbatov, V. T. Volkov, L. Vaccarini, and H. Bouchiat, *Phys. Rev. Lett.* **86**, 2416 (2001).

[49] R. A. Smith, B. S. Handy, and V. Ambegaokar, *Phys. Rev. B* **63**, 094513 (2001).

[50] P. Xiong, A. V. Herzog, and R. C. Dynes, *Phys. Rev. Lett*. **78**, 927 (1997).

[51] H. Miyazaki, Y. Takahide, A. Kanda, and Y. Ootuka, *Phys. Rev. Lett*. **89**, 197001 (2002).